\documentclass[a4paper,11pt]{article}

\usepackage{amsfonts,mathrsfs,mathtools,bm,amsmath,xcolor}
\usepackage[margin=25mm]{geometry}
\usepackage{enumitem}
\usepackage{natbib}
\usepackage{booktabs}
\usepackage{amssymb}

\newcommand{\norm}{\mathrm{N}}
\newcommand{\gam}{\mathrm{Gam}}
\newcommand{\E}{\mathnormal{E}}
\newcommand{\Var}{\text{Var}}
\newcommand{\Cov}{\text{Cov}}
\newcommand{\Cor}{\text{Cor}}
\newcommand{\diag}{\mathrm{diag}}
\newcommand{\vect}{\mathrm{vec}}
\def\T{{ \mathrm{\scriptscriptstyle T} }}

\renewcommand{\vec}[1]{\boldsymbol{#1}}
\newcommand{\vecn}[1]{\boldsymbol{#1}}
\newcommand{\matr}[1]{\mathnormal{#1}}
\newcommand{\matrn}[1]{\mathrm{#1}}
\newcommand{\VAR}[2]{\textsc{VAR}$_{#1}\text{(}{#2}\text{)}$} 
\newcommand{\VARMA}[3]{\textsc{VARMA}$_{#1}\text{(}{#2}\text{,}{#3}\text{)}$}

\definecolor{five1}{HTML}{F8766D}
\definecolor{five2}{HTML}{A3A500}
\definecolor{five3}{HTML}{00BF7D}
\definecolor{five4}{HTML}{00B0F6}
\definecolor{five5}{HTML}{E76BF3}

\providecommand{\keywords}[1]
{
  \small
  \textbf{\textit{Keywords---}} #1
}

\title{Enforcing stationarity through the prior in vector autoregressions}
\author{Sarah E. Heaps \\
\small Durham University, Durham, U.K.\\
\small Email: \texttt{sarah.e.heaps@durham.ac.uk}}
\date{}

\begin{document}

\maketitle

\begin{abstract}
Stationarity is a very common assumption in time series analysis. A vector autoregressive process is stationary if and only if the roots of its characteristic equation lie outside the unit circle, constraining the autoregressive coefficient matrices to lie in the stationary region. However, the stationary region has a highly complex geometry which impedes specification of a prior distribution. In this work, an unconstrained reparameterization of a stationary vector autoregression is presented. The new parameters are partial autocorrelation matrices, which are interpretable, and can be transformed bijectively to the space of unconstrained square matrices through a simple mapping of their singular values. This transformation preserves various structural forms of the partial autocorrelation matrices and readily facilitates specification of a prior. Properties of this prior are described along with an important special case which is exchangeable with respect to the order of the elements in the observation vector. Posterior inference and computation are described and implemented using Hamiltonian Monte Carlo via Stan. The prior and inferential procedures are illustrated with an application to a macroeconomic time series which highlights the benefits of enforcing stationarity and encouraging shrinkage towards a sensible parametric structure. Supplementary materials for this article are available in the ancillary files section.
\end{abstract}


\section{\label{sec:intro}Introduction}

Denote by $\{ \vec{y}_t \}$ a time series of equally spaced $m$-variate observations. A stochastic process is said to be strictly stationary if its properties are unaffected by a shift in the time origin and weakly stationary if the mean $\E(\vec{y}_t) = \vec{\mu}$ remains constant over time and the autocovariance function $\matr{\Gamma}_i = \Cov(\vec{y}_t, \vec{y}_{t+i}) = \E\{(\vec{y}_t - \vec{\mu}) (\vec{y}_{t+i} - \vec{\mu})^\T\}$ depends only on the lag $i$ ($i=0,1,\ldots$), with $\matr{\Gamma}_{-i} = \matr{\Gamma}_i^\T$. For Gaussian processes, the two are equivalent, and we simply refer to a process as being stationary. As raw time series often exhibit periodic variation and systematic changes in the mean, stationarity is generally assumed only for the residuals of a detrended series, the variables of a differenced process, or those latent components of a state space model that are believed to be mean-reverting. In these cases, stationarity prevents the predictive variance of the transformed process from growing without bound as the forecast horizon increases, moving either forwards or backwards in time. This is a very reasonable assumption in many applications, for instance where the inferential objective is long-term forecasting or characterizing the long-run behaviour of linear dynamic systems.

All stationary Gaussian processes can be approximated arbitrarily well by a finite-order, vector autoregressive moving average (VARMA) model \citep[][Chapter 12]{Neu16}. Although we focus on the more commonly used subclass of vector autoregressive (VAR) models, we discuss the extension to the more general case in Section~\ref{sec:discussion}. Without loss of generality, assume that the time series $\{ \vec{y}_t \}$ can be modelled as a zero-mean, order-$p$ vector autoregressive, or \VAR{m}{p}, process, $\vec{y}_t = \matr{\phi}_1 \vec{y}_{t-1} + \cdots + \matr{\phi}_p \vec{y}_{t-p} + \vec{\epsilon}_t$, in which the errors $\{ \vec{\epsilon}_t \}$ form a sequence of uncorrelated, zero-mean multivariate normal random vectors, $\vec{\epsilon}_t \sim \norm_m(\vecn{0}, \matr{\Sigma})$. The parameters of the model therefore comprise the autoregressive coefficient matrices $\matr{\phi}_i \in M_{m \times m}(\mathbb{R})$ ($i=1,\ldots,p$), and the error variance matrix $\matr{\Sigma} \in \mathcal{S}^+_m$, where $M_{m \times n}(V)$ denotes the set of $m \times n$ matrices with entries in $V$ and $\mathcal{S}^+_m$ denotes the set of $m \times m$ symmetric, positive definite matrices. Henceforth, we refer to the collection of $\matr{\phi}_i$ as $\matr{\Phi}$. \label{pg:rev1_1c}VAR processes are a widely used class of time-series model which find application in a broad range of fields, such as macroeconomics \citep[][]{KK09}, neuroscience \citep[][]{CGYHSV17}, and genomics \citep[][]{AW13}. In most cases, priors for $(\matr{\Phi}, \matr{\Sigma})$ are based on either the conjugate matrix normal inverse Wishart distribution \citep[][]{BGR10} or the semi-conjugate variant in which $\matr{\Phi}$ and $\matr{\Sigma}$ are independent \textit{a priori} \citep[][]{Kar13}. Variations include priors that fix the error variance at an estimate, such as the Minnesota prior \citep[][]{DLS84}, or hierarchical priors which are typically designed to allow data-informed shrinkage towards a sparse parameterization \citep[][]{GKM19}. Another class of priors arises in the context of graphical VARs where the idea is to learn the contemporaneous and lagged relationships over time \citep[][]{CV05,PC20}. In this case, a prior is first placed on the graph underpinning the process and then a (typically vague) prior is placed on the parameters that are unrestricted after conditioning on that graph. In this literature, although stationarity is often stated as an assumption to derive properties of the process, it is not enforced as a constraint.

It is common to write the \VAR{m}{p} process as $\vec{\epsilon}_t = (\matr{I}_m - \matr{\phi}_1 B - \cdots - \matr{\phi}_p B^p) \vec{y}_t = \matr{\phi}(B) \vec{y}_t$ where $B$ is the backshift operator, that is $B \vec{y}_t = \vec{y}_{t-1}$, $\matr{I}_m$ is the $m \times m$ identity matrix and $\matr{\phi}(u) = \matr{I}_m - \matr{\phi}_1 u - \cdots - \matr{\phi}_p u^p$, $u \in \mathbb{C}$, is termed the characteristic polynomial. The process is stationary if and only if all the roots of $\det \{ \matr{\phi}(u) \} = 0$ lie outside the unit circle. We refer to this subset of $M_{m \times m}(\mathbb{R})^p$ as the stationary region, denoted by $\mathcal{C}_{p,m}$.

When $m=1$ and $p=1$ or $p=2$, the stationary region is simple, with $\mathcal{C}_{1,1}$ representing the interval $(-1, 1)$, and $\mathcal{C}_{2,1}$ a triangle in the $(\phi_1, \phi_2)$-plane. However, increasing $m$ or $p$ further increases the complexity of the polynomial equation $\det \{ \matr{\phi}(u) \} = 0$ and hence the geometry of the stationary region. This causes two main problems for Bayesian inference. First, because there are no standard distributions on $\mathcal{C}_{p,m}$, it is not generally possible to directly specify a prior over this space which encodes genuine beliefs. Second, designing efficient Markov chain Monte Carlo (MCMC) samplers with state space constrained to $\mathcal{C}_{p,m}$ is very challenging. In the univariate ($m=1$) case, \citet{Chi93} addresses the latter difficulty by assigning a multivariate normal prior to $\matr{\Phi}$, truncated to the region $\mathcal{C}_{p,1}$. For $p>1$, the normalizing constant cannot be evaluated in closed form and so the prior density is known only up to proportionality. A sampler is described which updates $\matr{\Phi}$ in a block in which the proposal is the full conditional distribution when the constraints are ignored; proposals that fall outside of $\mathcal{C}_{p,1}$ are rejected in the Metropolis acceptance step. However, this will be efficient only if the proposal density is concentrated over $\mathcal{C}_{p,1}$. In the univariate case, \citet{Pic82} calculates the volume of the stationary region for the parameters of autoregressive (AR) models. In the special case where the autoregressive coefficient matrices $\matr{\phi}_i$ are diagonal, a trivial corollary of this result is that the volume of the stationary region is equal to $U_p^m$, in which $U_p = (M_1 M_3 \cdots M_{p-1})^2$ for $p$ even and $U_{p}=U_{p-1} M_{p}$ for $p$ odd, where $M_i = 2^i \left[ \{(i-1)/2\} ! \right]^2 / i!$.  This region becomes vanishingly small as $p$ increases and is likely to render an inferential scheme that is not tailored to the geometry of the problem highly inefficient. A natural way around both of these problems is to find an interpretable reparameterization of the \VAR{m}{p} model which maps $\matr{\Phi} \in \mathcal{C}_{p,m}$ to a space which has simpler constraints and allows prior specification to be carried out in a meaningful way.

As a result of their ubiquity in time series analysis, there is a large literature on reparameterizations of univariate AR models. \citet{BS73} establish a bijection between $\matr{\Phi} \in \mathcal{C}_{p,1}$ and the first $p$ partial autocorrelations $\vec{\rho} = (\rho_1, \ldots, \rho_p)$ of a stationary, $p$th order autoregression. \citet{Mon84} provides an alternative derivation of the mapping and explicit recursive formulae for its inverse. The new parameterization has the advantage that the partial autocorrelations $\vec{\rho}$ are interpretable and only constrained to lie in the Cartesian product space $(-1,1)^p$. \citet{MRGP96} and \citet{BKS96} present prior distributions for the partial autocorrelations and MCMC algorithms for computational inference. The latter suggest a uniform prior over $\mathcal{C}_{p,1}$ which induces a closed form for the density of $\vec{\rho}$ due to the analytical tractability of the Jacobian term. The former use spike-and-slab priors for the $\rho_i$ in which the slab is uniform over $(-1, 1)$ and the spike is an atom of probability at zero, which allows for uncertainty in the model order. \citet{MS92} and \citet{HW99} describe reparameterizations based on a representation of the characteristic equation in factorized form, that is, $\phi(u) = \prod_{i=1}^p (1 - \eta_i u)$. In this case, the condition for stationarity reduces to $| \eta_i | < 1$ ($i=1,\ldots,p$). \citet{HW99} also allow uncertainty in the model order $p$, and the balance of real and complex (reciprocal) roots $\eta_i$, by placing priors on the real roots and the moduli of the pairs of complex roots with atoms of probability at zero.

Extensions of these reparameterizations of univariate AR models to the general vector case, especially with a focus on prior specification, are surprisingly scarce in the literature. \citet{HP06} extend the ideas of \citet{HW99}, but only in the special case where the $\matr{\phi}_i$ are diagonal. \label{pg:rev1_5a}\citet{HSW13} describe a sparsity-inducing penalized maximum likelihood algorithm for frequentist model-fitting. However, it is not fully flexible, constraining inference to a subset of the stationary region where the spectral norm, rather than spectral radius, of the companion matrix is less than one. \citet{MVK78} generalize the results of \citet{BS73}; for every error variance matrix $\matr{\Sigma}$, a bijection is established between $\matr{\Phi} \in \mathcal{C}_{p,m}$ and the first $p$ partial autocorrelation matrices $\matr{P}_1, \ldots, \matr{P}_p$ of a stationary \VAR{m}{p} process. Denoting by $\mathcal{V}_m$ the subset of matrices in $M_{m \times m}(\mathbb{R})$ whose singular values are all less than one, in terms of the new parameters, the stationary region reduces to a simple Cartesian product space $(\matr{P}_1, \ldots, \matr{P}_p) \in \mathcal{V}_m^p$. \citet{AK86} build on the earlier work in \citet{AN79} by generalizing the construction of \citet{Mon84} and explicitly providing recursive formulae for the inverse mapping. A second bijective mapping between $(\matr{P}_1, \ldots, \matr{P}_p) \in \mathcal{V}_m^p$ and $(\matr{A}_1, \ldots, \matr{A}_p) \in M_{m \times m}(\mathbb{R})^p$ is described along with a maximum likelihood estimation procedure. Although the new parameters $\matr{A}_1, \ldots, \matr{A}_p$ are unconstrained, the reparameterization is not immediately amenable to Bayesian inference because the $\matr{A}_i$ are difficult to interpret. \citet{RML19} derive an alternative reparameterization of stationary and invertible VARMA models. The new parameters comprise a set of $p$ symmetric, positive definite matrices and a set of $p$ orthogonal matrices. From an inferential perspective, the space in which the new parameters lie is almost as problematic as the stationary region $\mathcal{C}_{p,m}$. As we discuss further in Section~\ref{subsubsec:rml}, prior specification is difficult because the orthogonal matrices lack a meaningful interpretation, whilst computational inference remains challenging because the constraints that define the space of orthogonal matrices complicate sampling by MCMC.

\label{pg:rev1_1d}In this paper we propose a different reparameterization and prior for the parameters of a stationary VAR process. Like in \citet{AK86}, the new parameters are constructed through two mappings: first to a set of (interpretable) partial autocorrelation matrices $\matr{P}_1, \ldots, \matr{P}_p$ and then to a set of unconstrained square matrices $\matr{A}_1, \ldots, \matr{A}_p$. Our reparameterization differs, however, by facilitating the preservation of certain symmetries in the first mapping and the interpretability of the partial autocorrelation matrices in the second mapping. This allows construction of prior distributions for $(\matr{A}_1, \ldots, \matr{A}_p)$ which encourage shrinkage towards meaningful parametric structures. In particular, we describe a prior that is exchangeable with respect to the order of the elements in the observation vector. This is likely to be a useful representation of prior beliefs in a variety of applications where the modeller does not have information, \textit{a priori}, to distinguish between the $m$ time series. We also present an inferential scheme that allows computational inference to be carried out using Euclidean Hamiltonian Monte Carlo, implemented by the probabilistic programming software Stan  \citep[][]{CGH17}. This facilitates straightforward extension and adaptation by users in the wide-variety of fields in which vector autoregressions are used. The main contributions of the paper are therefore two-fold: first, a reparameterization and prior for stationary VARs that allows the incorporation of meaningful initial beliefs; and second, routine implementation of computational inference using standard probabilistic programming software. 

\section[Reparameterization over the stationary region]{\label{sec:reparam}Reparameterization over the stationary region}

\subsection{\label{subsec:part1_mapping}Reparameterization via partial autocorrelation matrices}
\citet{AK86} establish a one-to-one correspondence between the parameters of a stationary \VAR{m}{p} process $( \matr{\Sigma}, \matr{\Phi} ) \in \mathcal{S}^+_m \times \mathcal{C}_{p,m}$ and the parameter set $\{ \matr{\Sigma}, (\matr{P}_1, \ldots, \matr{P}_p) \} \in \mathcal{S}^+_m \times \mathcal{V}_{m}^p$ in which $\matr{P}_s$ denotes the $s$th partial autocorrelation matrix. In essence, the $(s+1)$th partial autocorrelation matrix  $\matr{P}_{s+1}$ is a conditional cross-correlation matrix between $\vec{y}_{t+1}$ and $\vec{y}_{t-s}$ given $\vec{y}_t, \ldots, \vec{y}_{t-s+1}$. More precisely, the matrices $\matr{P}_1, \ldots, \matr{P}_p$ are defined as follows. For each $s=1,\ldots,p$ let
\begin{equation}\label{eq:forrevreg}
\vec{y}_{t+1} = \sum_{i=1}^s \matr{\phi}_{si} \vec{y}_{t-i+1} + \vec{\epsilon}_{s,t+1}, \quad \vec{y}_{t-s} = \sum_{i=1}^s \matr{\phi}^*_{si} \vec{y}_{t-s+i} + \vec{\epsilon}^*_{s,t-s}
\end{equation}
in which the $m \times m$ matrices $\matr{\phi}_{si}$ and $\matr{\phi}^*_{si}$ are the coefficients of the $i$th terms $\vec{y}_{t-i+1}$ and $\vec{y}_{t-s+i}$, respectively, in the conditional expectations $\E(\vec{y}_{t+1} \mid \vec{y}_{t}, \ldots, \vec{y}_{t-s+1})$ and $\E(\vec{y}_{t-s} \mid \vec{y}_{t-s+1}, \ldots, \vec{y}_{t})$. It follows that $\matr{\phi}_{pi}=\matr{\phi}_i$ ($i=1,\ldots,p$) and $\matr{\Sigma}_p=\matr{\Sigma}$. Equivalently, because the multivariate normal distribution is defined by its first two moments, the $\matr{\phi}_{si}$ and $\matr{\phi}^*_{si}$ are the values of the coefficients, say $\matr{\alpha}_i$ and $\matr{\alpha}_i^*$, in the autoregression of $\vec{y}_{t+1}$ on its $s$ predecessors or successors, respectively, that minimize the mean squared error $\E\{(\vec{y}_{t+1} - \sum_{i=1}^s \matr{\alpha}_{i} \vec{y}_{t-i+1})^\T (\vec{y}_{t+1} - \sum_{i=1}^s \matr{\alpha}_{i} \vec{y}_{t-i+1}) \}$ or $\E\{(\vec{y}_{t-s} - \sum_{i=1}^s \matr{\alpha}_{i}^* \vec{y}_{t-s+i})^\T (\vec{y}_{t-s} - \sum_{i=1}^s \matr{\alpha}_{i}^* \vec{y}_{t-s+i})\}$; see, for example, Chapter 3 of \citet{Chr91}. We define the corresponding conditional variances as $\matr{\Sigma}_s = \Var(\vec{\epsilon}_{s,t+1}) = \Var(\vec{y}_{t+1} \mid \vec{y}_{t}, \ldots, \vec{y}_{t-s+1})$ and $\matr{\Sigma}^*_s = \Var(\vec{\epsilon}^*_{s,t-s}) = \Var(\vec{y}_{t-s} \mid \vec{y}_{t-s+1}, \ldots, \vec{y}_{t})$ ($s=1,\ldots,p$) and let $\matr{\Sigma}_0 = \matr{\Sigma}^*_0 = \matr{\Gamma}_0$ where $\matr{\Gamma}_i = \Cov(\vec{y}_t, \vec{y}_{t+i}) = \E(\vec{y}_t \vec{y}_{t+i}^\T)$ is the $i$th autocovariance of $\vec{y}_t$. Now, express the conditional variance matrices through a matrix-square-root decomposition, $\matr{\Sigma}_s = \matr{S}_s \matr{S}_s^\T$ and $\matr{\Sigma}^*_s = \matr{S}^*_s \matr{S}^{* \, T}_s$. Although any unique matrix-square-root could be used, we take the symmetric matrix-square-root factorization in this paper, and so $\matr{S}_s=\matr{\Sigma}_s^{1/2}$ and $\matr{S}^*_s=\matr{\Sigma}_s^{* \, 1/2}$ are symmetric and positive definite. A different reparameterization, defined by \citet{AK86}, is based on the Cholesky factorization, in which case $\matr{S}_s$ and $\matr{S}^*_s$ are lower triangular. However, as we discuss in Section~\ref{subsubsec:exch_prior}, we prefer the symmetric factorization as it facilitates construction of a prior that is closed under orthogonal transformation of the observation vectors. Finally, let $\vec{z}_{0,t+1} = \matr{S}_0^{-1} \vec{y}_{t+1}$ and $\vec{z}^*_{0,t} = \matr{S}^{* \, -1}_0 \vec{y}_{t}$ be standardized versions of the forward and reverse time series and, for each $s=1,\ldots,p-1$, let $\vec{z}_{s,t+1} = \matr{S}_s^{-1} \vec{\epsilon}_{s,t+1}$ and $\vec{z}^*_{s,t-s} = \matr{S}^{* \, -1}_s \vec{\epsilon}^*_{s,t-s}$ be standardized versions of the forward and reverse error series. We can now define the partial autocorrelation matrix $\matr{P}_{s+1}$, $(s=0,\ldots,p-1)$, as 
\begin{equation*}
\matr{P}_{s+1} \! = \!  \Cov(\vec{z}_{s,t+1}, \vec{z}^*_{s,t-s}) \!  = \! \matr{S}_s^{-1} \Cov( \vec{y}_{t+1}, \vec{y}_{t-s} | \vec{y}_t, \ldots, \vec{y}_{t-s+1}) (\matr{S}^{* \, -1}_s)^\T \! = \!  \matr{S}_s^{-1} \matr{\phi}_{s+1,s+1} \matr{S}_s^{*}
\end{equation*}
in which the final equality is demonstrated in the Supplementary Materials. This simplifies to the well-known result $\rho_{s+1}=\phi_{s+1,s+1}$ in the univariate case.

The (recursive) mapping from $( \matr{\Sigma}, \matr{\Phi} ) \in \mathcal{S}^+_m \times \mathcal{C}_{p,m}$ to $\{ \matr{\Sigma}, (\matr{P}_1, \ldots, \matr{P}_p) \} \in \mathcal{S}^+_m \times \mathcal{V}_{m}^p$ and its inverse are described in the Appendix with proofs in the Supplementary Materials. \label{pg:rev2_1a}Use of symmetric matrix-square-roots, rather than Cholesky factors, complicates the reverse map and precludes use of Lemma 2.3 from \citet{AK86} to perform the calculation. This is remedied through a novel recursion which allows computation of the stationary variance matrix $\matr{\Gamma}_0$ from the new set of parameters. 

The constraints on the $p$-fold Cartesian product space $\mathcal{V}_m^p$ are substantially simpler than those on the stationary region $\mathcal{C}_{p,m}$. \label{pg:rev2_4}Indeed, for each $\matr{P}_s \in \mathcal{V}_m$, the constraint can be expressed as an inequality for the spectral norm, $\| \matr{P}_s \|_2 < 1$, and so it may be possible to apply the spherical augmentation technique \citep[][]{LS16}, designed for handling norm constrains, to sample from probability distributions over $\mathcal{V}_m$. However, though the partial autocorrelation matrices are a very natural set of quantities about which to elicit prior beliefs, there are no standard distributions defined over $\mathcal{V}_m$. Further, any distribution on $M_{m \times m}(\mathbb{R})$ that was simply truncated to $\mathcal{V}_m$ would have an intractable normalizing constant, making its properties difficult to assess. This makes direct specification of a meaningful prior over $\mathcal{V}_m^p$ untenable. We therefore choose to apply a second reparameterization which maps the partial autocorrelations to unconstrained Euclidean space.

\subsection{\label{subsec:part2}Reparameterization via unconstrained square matrices}

In the second reparameterization, each $\matr{P} \in \mathcal{V}_m$ is mapped to an unconstrained square matrix $\matr{A} \in M_{m \times m}(\mathbb{R})$. Generalizing the one-to-one mapping defined by \citet{AK86} so that it involves arbitrary matrix-square-roots, the forward mapping is defined as follows. Let
\begin{equation}\label{eq:PtoA}
\matr{B}^{-1} \matr{B}^{-1 \, T} = \matr{I}_m - \matr{P} \matr{P}^\T
\end{equation}
be a matrix-square-root factorization of $\matr{I}_m - \matr{P} \matr{P}^\T$. Then write $\matr{A} = \matr{B} \matr{P}$. Similarly, for the inverse mapping, let 
\begin{equation}\label{eq:AtoP}
\matr{B} \matr{B}^\T = \matr{I}_m + \matr{A} \matr{A}^\T
\end{equation}
be a matrix-square-root factorization of $\matr{I}_m + \matr{A} \matr{A}^\T$, then write $\matr{P} = \matr{B}^{-1} \matr{A}$.

\label{pg:rev1_1b}Although \citeauthor{AK86} define the mapping in terms of the Cholesky factorization, we, instead, propose use of symmetric matrix-square-roots because it gives the new parameters a more natural interpretation. Specifically, denote the singular value decomposition of $\matr{P}$ by $\matr{P}=\matr{U} \matr{R} \matr{V}^\T$ in which the diagonal matrix $\matr{R} = \diag(r_1, \ldots, r_m)$ contains the $m$ singular values. These satisfy $1 > r_1 \ge r_2 \ge \cdots \ge r_m \ge 0$. It is straightforward to show that if symmetric square-roots are used, the corresponding singular value decomposition of $\matr{A}$ is $\matr{A} = \matr{U} \matr{\tilde{R}} \matr{V}^\T$ in which $\matr{\tilde{R}} = (\matr{I}_m - \matr{R}^2)^{-1/2} \matr{R}$ is a diagonal matrix whose $i$th diagonal element is $\tilde{r}_i = r_i / (1 - r_i^2)^{1/2} \ge 0$. It follows that $\matr{P}$ and $\matr{A}$ share the same singular vectors and that the singular values of $\matr{A}$ are a strictly increasing function of the singular values of $\matr{P}$. Clearly, the same functional relationship connects their spectral norms, $r_1 = \| \matr{P} \|_2$ and $\tilde{r}_1 = \| \matr{A} \|_2$ and so the relative sizes of $\| \matr{A}_s \|_2$ across lags $s$ can be interpreted as indicating the relative magnitudes of the partial autocorrelations at each lag. If the symmetric matrix-square-root factorization is used in~\eqref{eq:PtoA} and~\eqref{eq:AtoP} we can therefore think of this second reparameterization as an orientation-preserving transformation from $\matr{P}$ to $\matr{A}$ which simply maps the singular values from $[0,1)$ to the positive real line. As a direct consequence, the transformation preserves various meaningful parametric forms whose singular vectors only depend on the elements of the matrix through the requisite ordering of the singular values. Specific examples are detailed below.

From the singular value decomposition of a diagonal matrix it is clear that $\matr{P} \in \mathcal{V}_m$ is diagonal, with $j$th diagonal element $p_{jj}=p_j \in [0, 1)$, if and only if $\matr{A}$ is diagonal, with $j$th diagonal element $a_{jj}=p_j / (1 - p_{j}^2)^{1/2} \in \mathbb{R}$ ($j=1,\ldots,m$). A corollary is that $\matr{P}$ is a scaled identity matrix if and only if the same is true of $\matr{A}$. As a special case, when this scaling constant is equal to zero, $\matr{P} = \matrn{0}_m$ if and only if $\matr{A} = \matrn{0}_m$, where $\matrn{0}_m$ denotes an $m \times m$ matrix of zeros. This is a particularly useful theoretical result because it follows directly from the definition of the partial autocorrelation matrix that for $k<p$, $\matr{P}_k \ne \matrn{0}_m$ and $\matr{P}_{k+s} = \matrn{0}_m$ for $s=1,\ldots,p-k$ if and only if $\matr{\phi}_k \ne \matrn{0}_m$ and $\matr{\phi}_{k+s} = \matrn{0}_m$ for $s=1,\ldots,p-k$. The order of the VAR model is therefore $k < p$ if and only if $\matr{A}_k \ne \matrn{0}_m$ and $\matr{A}_{k+s} = \matrn{0}_m$ for $s=1,\ldots,p-k$. We return to this point in Section~\ref{sec:discussion}.

Now, consider a two-parameter exchangeable matrix defined by $(b - c) \matr{I}_m + c \matr{J}_m$ where $\matr{J}_m = \vecn{1}_m \vecn{1}_m^\T$ and $\vecn{1}_m$ is an $m$-vector of 1s. This is the most general form for a $m \times m$ square matrix which is invariant under a common permutation of the rows and columns. It is straightforward to show that a matrix of this form has a singular value decomposition whose singular vectors depend only on $m$ and the ordering of the singular values, $|b - c|$ and $| b + (m-1)c |$, which have multiplicity $m-1$ and $1$, respectively. It follows that $\matr{P} \in \mathcal{V}_m$ is a two-parameter exchangeable matrix if and only if the same is true of $\matr{A}$. The necessary and sufficient condition for $\matr{P} = (p_1 - p_2) \matr{I}_m + p_2 \matr{J}_m$ to lie in $\mathcal{V}_m$ can be expressed as $|p_{1}'| < \surd 2 / 2$ and $|p_{2}'| < \surd 2 / m$ where
\begin{equation}\label{eq:twoparam1and2}
p_{1}' = (p_{1}-p_{2}) \surd 2 / 2, \quad  p_{2}' = \left\{ p_{1} + (m-1) p_{2} \right\} \surd 2 / m.
\end{equation}
It is then straightforward to show that the corresponding unconstrained square matrix $\matr{A} = (a_{1} - a_{2}) \matr{I}_m + a_{2} \matr{J}_m$, with $a_{1}, a_{2} \in \mathbb{R}$, is such that
\begin{equation}\label{eq:twoparam3and4}
a_{i} = \frac{\left\{\surd 2 m p_{1}'\left(2-m^2 p_{2}'^2\right)^{1/2}\right\}(2-i) - \surd 2 p_{1}'\left(2-m^2 p_{2}'^2\right)^{1/2} + m p_{2}'\left(1-2p_{1}'^2\right)^{1/2}}{m\left\{\left(2-m^2 p_{2}'^2\right)\left(1-2p_{1}'^2\right)\right\}^{1/2}}.
\end{equation}

\section{\label{sec:prior}Prior distributions over the unconstrained space}

\subsection{\label{subsec:general_form}General form}
Let $\vect(\cdot)$ denote the vectorization operator. Conditional on a set of unknown hyperparameters, we construct a prior distribution with joint density
\begin{equation}\label{eq:prior}
\pi(\matr{\Sigma}, \matr{A}_1, \ldots, \matr{A}_p) = \pi(\matr{\Sigma}) \prod_{s=1}^p \pi\{ \vect(\matr{A}_s^\T) \}
\end{equation}
in which $\matr{\Sigma}$ is assigned a distribution over $\mathcal{S}^+_m$ and $\vect(\matr{A}_s^\T)$ ($s=1,\ldots,p$), is assigned a multivariate normal distribution. Predominantly, our focus in this paper is  specification of a prior for the latter. 

\subsection{\label{subsubsec:exch_prior}Exchangeable prior distribution}

Consider any $m \times m$ orthogonal matrix $\matr{H}$. Assuming that symmetric matrix-square-roots are used in both parts of the reparameterization of a stationary \VAR{m}{p} model for $\vec{y}_t$ ($t=1,2,\ldots$), we show in the Supplementary Materials that the parameters of the stationary \VAR{m}{p} model for $\vec{\tilde{y}}_t = \matr{H} \vec{y}_t$ are $\matr{\tilde{\Sigma}}=\matr{H} \matr{\Sigma} \matr{H}^\T$ and $\matr{\tilde{A}}_s=\matr{H} \matr{A}_s \matr{H}^\T$ ($s=1,\ldots,p$). It follows that if $\matr{\Sigma}$ and $\matr{A}_s$ ($s=1,\ldots,p$) are assigned a prior from the same distributional family as that of $\matr{H} \matr{\Sigma} \matr{H}^\T$ and $\matr{H} \matr{A}_s \matr{H}^\T$  ($s=1,\ldots,p$), then the prior induced for $( \matr{\Sigma}, \matr{\Phi} )$ over $\mathcal{S}^+_m \times \mathcal{C}_{p,m}$ will be closed under orthogonal transformation of the observation vectors. Priors for $\matr{\Sigma}$ over $\mathcal{S}^+_m$ possessing this closure property include the inverse Wishart distribution and the multivariate normal distribution for the matrix logarithm, $\log \matr{\Sigma}$ \citep[][]{LH92}. For $\vect(\matr{A}_i^\T) \in \mathbb{R}^{m^2}$, a multivariate normal prior meets this requirement. \label{pg:rev1_1a}It is important to note that if Cholesky factors were used in the first and second part of the reparameterization, such a prior would not be available because the partial autocorrelation matrices $\matr{\tilde{P}}_s$ and associated unconstrained $\matr{\tilde{A}}_s$ would not be orthogonal similarity transformations of $\matr{P}_s$ and $\matr{A}_s$.

In the analysis of multivariate stochastic processes, we often do not have information, \textit{a priori}, to distinguish between the $m$ components of $\vec{y}_t$. In this case it is reasonable to assign $( \matr{\Sigma}, \matr{\Phi} )$ a prior which is exchangeable with respect to the ordering of the elements in the observation vector. Because of its closure under orthogonal transformation, we can obtain an exchangeable prior by restricting our attention to distributions for $\matr{\Sigma}$ and $\matr{A}_s$ that would be exactly the same as the distributions for $\matr{H} \matr{\Sigma} \matr{H}^\T$ and $\matr{H} \matr{A}_s \matr{H}^\T$ for any permutation matrix $\matr{H}$. That is, distributions for $\matr{\Sigma}$ and $\matr{A}_s$ which are invariant under a common permutation of the rows and columns. Given certain choices of their hyperparameters, the (conjugate) inverse Wishart distribution and the multivariate normal distribution for the matrix-logarithm can yield an exchangeable prior for $\matr{\Sigma}$. For instance, we could assign $\matr{\Sigma}$ an inverse Wishart prior with a (positive definite) two-parameter exchangeable scale matrix.

Now, suppose we wish to assign an exchangeable prior to $\matr{A}_s=(a_{s,ij})$ ($s=1,\ldots,p$). Given the potential for the model to contain a very large number of parameters, suppose further that we want to specify a prior that allows borrowing of strength between the diagonal elements and between the off-diagonal elements of each $\matr{A}_s$. To this end, we can adopt a prior in which the diagonal and off-diagonal elements are given hierarchical distributions. At the top-level, we choose
\begingroup
\allowdisplaybreaks
\begin{alignat}{2}
a_{s,ii} \mid \mu_{s1}, \omega_{s1} &\sim \norm(\mu_{s1}, \omega_{s1}^{-1}),& \quad &\text{($i=1,\ldots,m$),}\label{eq:expriortop1}\\
a_{s,ij} \mid \mu_{s2}, \omega_{s2} &\sim \norm(\mu_{s2}, \omega_{s2}^{-1}),& \quad &\text{($i,j=1,\ldots,m$ with $i \ne j$).}\label{eq:expriortop2}
\end{alignat}
\endgroup
The mean and precision at the bottom level of the hierarchy can then be assigned priors on $\mathbb{R}$ and $\mathbb{R}^+$, such as
\begin{equation}
\mu_{si} \sim \norm(e_{si}, f_{si}^2), \quad \omega_{si} \sim \gam(g_{si}, h_{si}), \quad (i=1,2).\label{eq:expriorbot12}
\end{equation}
Marginally, $\E(a_{s,ii})=e_{s1}$, $\Var(a_{s,ii})=f_{s1}^2 + h_{s1} / (g_{s1} - 1)$ (for $g_{s1} > 1$) and $\Cor(a_{s,ii}, a_{s,jj})=f_{s1}^2(g_{s1} - 1) / \{ f_{s1}^2(g_{s1} - 1) + h_{s1} \}$ with similar expressions for the moments of the off-diagonal elements. Therefore, given specifications for the common diagonal elements and off-diagonal elements in $\matr{P}_s$, one can calculate corresponding values for the common diagonal and off-diagonal elements in $\matr{A}_s$ through~\eqref{eq:twoparam1and2}--\eqref{eq:twoparam3and4}. These can be taken as values for $e_{s1}=\E(a_{s,ii})$ and $e_{s2}=\E(a_{s,ij})$. Uncertainty in these central values, and the proportion of this which is shared among all the diagonal or all the off-diagonal elements, can be reflected through choices of the other hyperparameters. Clearly, specifications which make the marginal variances small and the marginal correlations large will shrink the posterior so that it is more concentrated over the space of two-parameter exchangeable structures. 

\subsection{\label{subsec:diag_prior}Prior distribution centred on a diagonal matrix}

Let $\vec{y}_{i:j} = (\vec{y}_i^\T, \ldots, \vec{y}_j^\T)^\T$. For each $j=1,\ldots,m$, suppose it is believed \textit{a priori} that once $\vec{y}_{t:(t-s+1)}$ is known, $y_{t-s,j}$ is the only element in $\vec{y}_{t-s}$ which provides further information about $y_{t+1,j}$. This is tantamount to a conjecture that the partial autocorrelation matrix $\matr{P}_s$, and hence $\matr{A}_s$, is diagonal; see Section~\ref{subsec:part2}. To represent this belief we choose a prior in which \mbox{$(a_{s,11}, \ldots, a_{s,mm})^\T \sim \norm_m(\vec{e}_s, \matr{F}_s)$} for the diagonal elements. Alternatively, if there was nothing in our prior beliefs to distinguish among the diagonal elements, and we wanted to allow borrowing of strength between them, we might adopt the hierarchical prior in~\eqref{eq:expriortop1} and \eqref{eq:expriorbot12} when $i=1$. For the off-diagonal elements $a_{s,ij}$ ($i \ne j$), we can centre our prior around zero, whilst allowing the data to influence the degree of shrinkage towards zero, by adopting a special case of the hierarchical prior defined by~\eqref{eq:expriortop2} and \eqref{eq:expriorbot12} in which the distribution for $\mu_{s2}$ is a point mass at zero. Alternatively, as we discuss further in Section~\ref{subsubsec:sparse}, the off-diagonal elements could be assigned a sparsity-inducing prior.

\subsection{\label{subsubsec:sparse}Prior distribution encouraging sparse $\matr{A}_s$ matrices}

A \VAR{m}{p} model is highly parameterized with $O(m^2)$ parameters. Indeed, particularly when $m$ is large, it is entirely plausible for there to be fewer observations in the data than there are parameters in the model. In a Bayesian analysis, this can lead to a diffuse posterior distribution for $( \matr{\Sigma}, \matr{\Phi} )$, making predictive distributions imprecise and complicating model interpretation. When stationarity is not enforced, this issue is often addressed by inducing sparsity amongst the elements of the (unconstrained) autoregressive coefficient matrices $\matr{\phi}_1, \ldots, \matr{\phi}_p$, either through graphical modelling or zero-mean shrinkage priors \citep{GSN08,BCR19}. The non-zero structure can then be associated with a directed graph representing a network of interactions because a zero in position $(i,j)$ of $\matr{\phi}_s$ implies conditional independence between $y_{t,i}$ and $y_{t-s,j}$ given $(\vec{y}_{t-1}, \ldots, \vec{y}_{t-s,-j}, \ldots, \vec{y}_{t-p})$ where $\vec{y}_{t-s,-j}=(y_{t-s,1}, \ldots, y_{t-s,j-1}, y_{t-s,j+1}, \ldots, y_{t-s,m})^\T$. In principle, sparsity-inducing priors could also be chosen for the elements of $\matr{A}_1, \ldots, \matr{A}_p$ in our parameterization, where stationarity is enforced. \label{pg:rev1_3}Among the partial autocorrelation matrices, a zero in position $(i,j)$ of $\matr{P}_s$ implies conditional independence between $y_{t,i}$ and $y_{t-s,j}$ given $\vec{y}_{(t-1):(t-s+1)}$. Transforming from $\matr{P}_s$ to $\matr{A}_s$, an individual zero in position $(i,j)$ of $\matr{P}_s$ does not give an individual zero in position $(i,j)$ of $\matr{A}_s$ or vice versa and so sparsity in $\matr{A}_s$ does not have a clear structural interpretation as it would for $\matr{P}_s$ or $\matr{\phi}_s$. However, the overall size of $\matr{A}_s$ is strongly linked to the overall size of $\matr{P}_s$ through the strictly increasing relationship that connects their spectral norms; see Section~\ref{subsec:part2}. Therefore, although the justification for a sparsity-inducing prior for the $\matr{A}_s$ is weaker from an explanatory perspective, there is still an argument for their use as a means of regularizing the variance of predictive distributions. 

A sparsity-inducing, zero-mean scale-mixture of normals prior would take the form $a_{s,ii} \mid \psi_{s,ii} \sim \norm(0, \psi_{s,ii})$ with $\psi_{s,ii} \sim \mathcal{F}_1$, independently for $i=1,\ldots,m$, and $a_{s,ij} \mid \psi_{s,ij} \sim \norm(0, \psi_{s,ij})$ with $\psi_{s,ij} \sim \mathcal{F}_2$, independently for $i,j=1,\ldots,m$ with $i \ne j$. Here $\mathcal{F}_1$ and $\mathcal{F}_2$ are mixing distributions which can either be discrete, as in spike-and-slab priors \citep[][]{MB88,GM93}, or continuous, as in the horseshoe \citep[][]{CPS10,PV17}.

\subsection{\label{subsubsec:rml}Vague prior via the parameterization of \citet{RML19}}

Denote by $\mathcal{O}(m)$ the space of $m \times m$ orthogonal matrices. \citet{RML19} establish a bijective mapping between the parameters of a stationary \VAR{m}{p} process $( \matr{\Sigma}, \matr{\Phi} ) \in \mathcal{S}^+_m \times \mathcal{C}_{p,m}$ and the parameter set $\{ \matr{\Sigma}, (\matr{V}_1, \ldots, \matr{V}_p), (\matr{Q}_1, \ldots, \matr{Q}_p) \} \in \mathcal{S}^+_m \times \mathcal{S}^{+ \, p}_{m} \times \mathcal{O}(m)^p$ for any fixed choice of a pseudo error variance matrix $\matr{M} \in \mathcal{S}^+_m$. It relies on a characterization in terms of positive definite block Toeplitz matrices, like that describing the variance of the joint stationary distribution of $p$ consecutive time points. \label{pg:rev2_1b}For the special case when $\matr{M}=\matr{\Sigma}$, we show in the Supplementary Materials that $\matr{V}_i$ represents the difference in conditional variances, $\matr{V}_{i} = \Var(\vec{y}_t \mid \vec{y}_{t-1}, \ldots, \vec{y}_{t-i+1}) - \Var(\vec{y}_t \mid \vec{y}_{t-1}, \ldots, \vec{y}_{t-i})$ ($i=1,\ldots,p$), and the orthogonal matrix $\matr{Q}_i$ arises from the polar decomposition of an affine transformation of the partial autocorrelation matrix $\matr{P}_i$. Unfortunately, this makes the orthogonal matrices difficult to interpret which impedes specification of a meaningful prior. \label{pg:rev1_2a}Moreover, computational inference is challenging because the constraints that define $\mathcal{O}(m)$ complicate sampling by MCMC. For example, attempts to provide a general reparameterization of $\matr{Q} \in \mathcal{O}(m)$ in terms of unconstrained parameters, such as the Givens representation \citep[][]{PJMAP19} or Cayley transform \citep[][]{JHD20}, are typically frustrated by the pathological effects of mapping between two topologically distinct spaces. \label{pg:rev1_2c}Indeed, as explained in the Supplementary Materials, the modified Cayley transform suggested by \citeauthor{RML19} is not bijective, \label{pg:revrev_rml3}which makes the posterior of their real-valued parameterization multimodal and can cause inefficient MCMC simulation. These geometric problems can be avoided by using Geodesic Monte Carlo \citep[][]{BG13} which is able to sample efficiently from $\mathcal{O}(m)$ by tailoring the Hamiltonian Monte Carlo method to embedded manifolds. However, there is currently no methodology for automatic tuning of its parameters, nor any modular software for implementation. This makes it difficult to put into practice through bespoke MCMC programs and impossible to implement using probabilistic programming software. Fortunately, in the Supplementary Materials we present a simple reparameterization of the parameter set $\{ (\matr{V}_1, \ldots, \matr{V}_p), (\matr{Q}_1, \ldots, \matr{Q}_p) \}$ in terms of $p$ unconstrained square matrices which circumvents the sampling issue. \label{pg:revrev_rml2}We show that this is equivalent to assigning independent uniform distributions over $\mathcal{O}(m)$ to the $\matr{Q}_s$ and independent Wishart distributions to the $\matr{V}_s$, with $m$ degrees of freedom and identity scale, and so it can be regarded as a vague, stationary prior distribution. This might be attractive to some modellers as a default choice of prior.

\subsection{\label{subsec:graphs}Choice of prior variance}

Using the simple example of a \VAR{2}{1} model, we show in the Supplementary Materials that the prior for the partial autocorrelation matrix $\matr{P}_1$ can become multimodal when the prior variance for the elements of the unconstrained square matrix $\matr{A}_1$ becomes too large. For most problems, a multimodal prior for a partial autocorrelation matrix $\matr{P}$ is unlikely to be representative of prior beliefs. To avoid this, care is clearly needed in the choice of prior variance for the elements of the unconstrained square matrices.

As discussed in Section~\ref{subsec:part2}, a partial autocorrelation matrix $\matr{P}$ is constructed from the corresponding unconstrained matrix $\matr{A}=(a_{ij})$ through a simple mapping of its singular values from the positive real line to the unit interval. It is reasonable, therefore, to conjecture that the multimodality that can occur in the prior for the partial autocorrelations, but not in the multivariate normal prior for the unconstrained matrices, arises through this mapping of the singular values. For any $m$, under the simple prior $a_{ij} \sim \norm(0, s^2)$ $(i,j=1,\ldots,m)$, we show in the Supplementary Materials that the singular values and right and left singular vectors of $\matr{P}$ are independent \textit{a priori}. Moreover, we show that the singular vectors are distributed as independent, normalized Haar measures, and derive an analytic expression for the joint prior density of the singular values. Arguing that multimodality arises when the singular values have a local maximum in the interior of their parameter space, we find the smallest prior standard deviation $s$ of the $a_{ij}$ at which this occurs for various values of $m$. The conclusion is that a prior standard deviation of $s=1$ should prevent multimodality for $m \ge 5$. Guidance on an upper limit for $s$ when $m=1,\ldots,4$ can be found in the Supplementary Materials. 

\section[Posterior inference through MCMC]{\label{sec:posterior}Posterior inference through MCMC}

Consider observations $\vec{y}_{1:n}$ modelled as realizations from a stationary \VAR{m}{p} process. The likelihood function can be expressed as
\begin{equation*}
p(\vec{y}_{1:n} \mid \matr{\Sigma}, \matr{\Phi}) = p(\vec{y}_{1:p} \mid \matr{\Sigma}, \matr{\Phi})  \prod_{t=p+1}^n p(\vec{y}_t \mid \vec{y}_{(t-p):(t-1)}, \matr{\Sigma}, \matr{\Phi})
\end{equation*}
in which $\vec{Y}_t \mid \vec{y}_{(t-p):(t-1)}, \matr{\Sigma}, \matr{\Phi} \sim \norm_{m}\left(\sum_{i=1}^p \matr{\phi}_i \vec{y}_{t-i} \, , \, \matr{\Sigma}\right)$ 
and the initial distribution is $(\vec{Y}_1^\T, \ldots, \vec{Y}_p^\T)^\T \mid \matr{\Sigma}, \matr{\Phi} \sim \norm_{mp}(\vecn{0}, \matr{G})$
where $\matr{G}$ is a positive definite block Toeplitz matrix with $\matr{\Gamma}_{j-i}$ as the block in rows $\{m(i-1)+1\}$ to $mi$ and columns $\{m(j-1)+1\}$ to $mj$ ($i,j=1,\ldots,p$) and $\matr{\Gamma}_{-k}=\matr{\Gamma}_k^\T$ ($k=1,\ldots,p-1$). For the purposes of evaluating the likelihood, the stationary variance $\matr{\Gamma}_0$ and covariances $\matr{\Gamma}_1, \ldots, \matr{\Gamma}_{p-1}$ are available as a by-product of the reverse mapping detailed in the Appendix.

Treating the likelihood as a function of the new parameters and combining it with the prior~\eqref{eq:prior} through Bayes theorem yields the posterior distribution as
\begin{equation*}
\pi(\matr{\Sigma}, \matr{A}_1, \ldots, \matr{A}_p, \vec{\vartheta} \mid \vec{y}_{1:n}) \propto \pi(\matr{\Sigma}) \pi(\vec{\vartheta}) \prod_{i=1}^p \pi\{ \vect(\matr{A}_i^\T) \mid \vec{\vartheta} \} p(\vec{y}_{1:n} \mid \matr{\Sigma}, \matr{A}_1, \ldots, \matr{A}_p),
\end{equation*}
in which $\vec{\vartheta}$ denotes any unknown hyperparameters in the prior for the $\matr{A}_s$. As each of the autoregressive coefficients $\matr{\phi}_s$ is a complicated function of the complete set of new parameters, $\{  \matr{\Sigma}, (\matr{A}_1, \ldots, \matr{A}_p) \}$, this posterior distribution neither has a standard form, nor admits any simple factorization that would arise from conditional independence amongst $\{  \matr{\Sigma}, (\matr{A}_1, \ldots, \matr{A}_p) \}$. As a consequence, it is ill-suited to MCMC methods that are based on Gibbs sampling. Indeed, our experience suggests that Metropolis-within-Gibbs samplers, which iterate through random-walk updates of $(\matr{\Sigma} \mid \matr{A}_1, \ldots, \matr{A}_p)$, $(\matr{A}_1 \mid \matr{A}_2, \ldots, \matr{A}_p, \matr{\Sigma})$, $\ldots$, $(\matr{A}_p \mid \matr{A}_1, \ldots, \matr{A}_{p-1}, \matr{\Sigma})$, perform very poorly as soon as $p>1$. It is therefore beneficial to use a sampler such as Hamiltonian Monte Carlo (HMC) \citep{GC11,Nea11} which uses information on the slope of the logarithm of the posterior density to generate global proposals that update all parameters simultaneously. We use \texttt{rstan} \citep[][]{Sta20}, the R interface to the Stan software, to implement the HMC algorithm. Stan requires users to write a program in the probabilistic Stan modelling language, the role of which is to provide instructions for computing the logarithm of the kernel of the posterior density function. The Stan software then automatically sets up a Markov chain simulation to sample from the resulting posterior. This includes calculation of the gradient of the logarithm of the posterior density, random initialization of the chains, and the tuning of the sampler.

\section{\label{sec:application}Application}

In this section, we illustrate use of our exchangeable prior and inferential methods by applying them to a quarterly time series of US macroeconomic data. The complete data set comprises measurements on 168 variables, running from quarter 1 of 1959 to quarter 4 of 2007. The variables are transformed to stationarity by differencing, sometimes after applying a log transformation, and then standardized, so it is reasonable to model the data as arising from a zero mean stationary process. A full description of the data and transformations can be found in \citet{Koo13}. Following analyses of the same data in \citet{Koo13} and \citet{KK09}, interest lies primarily in forecasting the first three variables: real GDP, the consumer price index, and an interest rate (Federal funds). Like these original analyses, we consider a small \VAR{3}{4} model and two larger \VAR{10}{4} and \VAR{20}{4} models, where the original three variables are supplemented by an additional seven, then a further ten, which are thought to have forecasting value. A list of these variables can be found in the Supplementary Materials. In order to assess the forecasting properties of the models, we fitted the models using data $\vec{y}_{1:n}$ where $n=156$ and held back the last 40 observations $\vec{y}_{(n+1):(n+40)}$ in all analyses. This allowed us to base our measures of forecasting performance on the posterior predictive distribution of the held-back data.

When it is reasonable to assume that a process is stationary, one of the problems of fitting an unconstrained VAR model is that some posterior mass often lies outside the stationary region. Typically, this is due to a combination of model misspecification and epistemic uncertainty in the parameter values that cannot be resolved by the data that have been observed. Therefore, to demonstrate the practical benefits of using our stationary prior, we additionally consider two commonly used priors that do not constrain inference to the stationary region. Specifically, we compare (i) the exchangeable, stationary prior from Section~\ref{subsubsec:exch_prior} to: (ii) a Minnesota prior; (iii) a semi-conjugate prior, which takes the form $\pi(\matr{\Phi}, \matr{\Sigma}) = \pi(\matr{\Phi}) \pi(\matr{\Sigma})$, where $\vect(\matr{\phi}_k) \sim \norm(\vec{u}_{k3}, \matr{W}_{k3})$ independently for $k=1,\ldots,4$ and $\matr{\Sigma} \sim \mathrm{IW}(m + 4, \vec{I}_m)$. Clearly, for the latter two analyses, the time series cannot be initialized at the stationary distribution, and so we simply condition on the first $p=4$ observations in the time series. For the exchangeable, stationary prior, we choose correlations of 0.7 between diagonal and between off-diagonal elements to facilitate borrowing of strength between elements. Under the Minnesota prior, the autoregressive coefficients are assigned a multivariate normal distribution, $\vect(\matr{\phi}_k) \sim \norm(\vec{u}_{k2}, \matr{W}_{k2})$ independently for $k=1,\ldots,4$, and the error variance matrix $\matr{\Sigma}$ is replaced by an estimate $\hat{\matr{\Sigma}} = \mathrm{diag}(s_1^2, \ldots, s_m^2)$, where $s_j^2$ is the ordinary least squares estimate of the error variance in the (univariate) autoregression for variable $j$. The prior mean $\vec{u}_{k2}$ is chosen so that $\E(\phi_{k,ij})=0$ if $i \ne j$ and the prior variance $\matr{W}_{k2}$ is taken to be diagonal; generally different variances are chosen for the $\phi_{k,ii}$ and the $\phi_{k,ij}$ for $i \ne j$ and the variances decrease as the lag $k$ increases. The idea is to encourage shrinkage towards a simple set of low order univariate AR models with the objective of reducing the epistemic component of the forecast variance. The complete prior specifications are provided in the Supplementary Materials. 

Even when inference is constrained to the stationary region, without appropriate borrowing of strength between parameters, posterior predictive distributions can still be overly diffuse, especially when $m$ is large. We therefore demonstrate the useful regularization effect that our exchangeable prior can provide by considering two further analyses where stationarity is guaranteed, but where there is no shrinkage towards a sensible parametric structure. In particular, we consider: (iv) the modified reparameterization of \citet{RML19}, and its associated vague prior, described in Section~\ref{subsubsec:rml}; (v) a frequentist, maximum likelihood analysis using the partial autocorrelation reparameterization of \citet{AK86} that is based on Cholesky factors rather than symmetric matrix-square-roots.

To fit the model under the exchangeable, stationary prior, the semi-conjugate prior and the vague, stationary prior we used HMC implemented by Stan. For all three data sets and all three priors, we used the \texttt{rstan} interface to the Stan software to run four chains, initialized at different starting points, for 2000 iterations, half of which were discarded as burn-in. \label{pg:rev2_7}The usual graphical and numerical diagnostics gave no evidence of any lack of convergence and, after pooling the chains, the effective sample size was at least 1262 for every parameter. The Minnesota prior is conjugate and so the posterior distribution can be computed analytically; see, for example, Chapter 2 of \citet{KK09} for its closed form. We generated 4000 independent draws from this distribution to allow calculation of the sample-based statistics described below. Model-fitting by maximum likelihood was also carried out using Stan which implements numerical maximization of the log-likelihood function by the quasi-Newton algorithm L-BFGS. We note that when $m=20$ the algorithm repeatedly failed to converge when initialized randomly and convergence was only achieved after initializing at the posterior mean deduced from the analysis under the exchangeable, stationary prior. The four Stan programs are given in the Supplementary Materials.

\begin{figure}[!t]
\begin{center}
\includegraphics[width=0.75\textwidth]{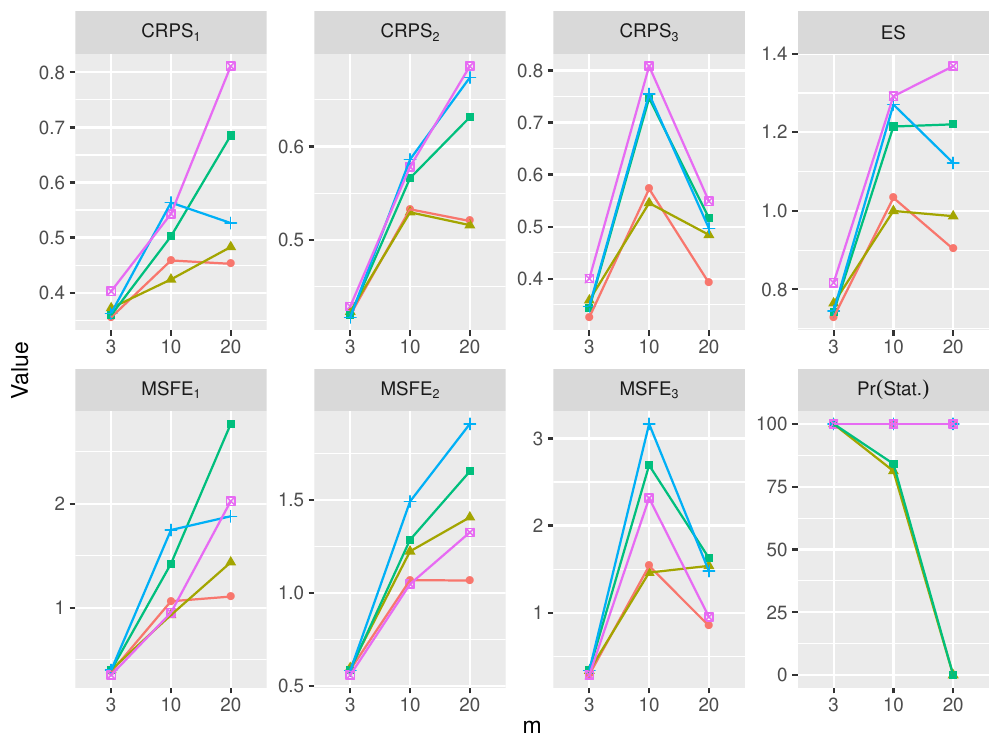}
\end{center}
\caption{For each value of $m$ and each prior: one-step-ahead CRPS for variable $k$ ($\mathrm{CRPS}_k$); one-step-ahead ES; posterior mean one-step-ahead MSFE for variable $k$ ($\mathrm{MSFE}_k$); posterior probability that $\matr{\Phi} \in \mathcal{C}_{4,m}$ ($\text{Pr}(\text{Stat.})$). The priors are: exchangeable and stationary (\textcolor{five1}{$\bullet$}); Minnesota (\textcolor{five2}{$\blacktriangle$}); semi-conjugate (\textbf{\textcolor{five3}{\scriptsize{$\blacksquare$}}}); vague and stationary prior (\textcolor{five4}{$+$}). Also shown are analogous statistics based on the stationary MLE \citep[][]{AK86} (\textcolor{five5}{\scriptsize{$\boxtimes$}}).\label{fig:scoring_rules}}
\end{figure}

To assess the forecasting performance of the various model-prior combinations we consider a variety of forecast horizons: $h=1,2,4,8$, ranging from short-term, one-quarter-ahead forecasting ($h=1$) to longer-term, two-year-ahead forecasting ($h=8$). For the comparison of $h$-step-ahead predictions, we use a number of proper scoring rules \citep[][]{GR07}, and the posterior for the empirical mean squared forecast error (MSFE) for each variable of interest. The MSFE is designed to measure the accuracy and precision of point forecasts, being based on the mean squared deviation between $\vec{y}_t$ and its expectation given $\vec{y}_{1:(t-h)}$ across the hold-out period, $t=n+h,\ldots,n+40$. Proper scoring rules compare the whole forecast distribution with the observation that arises; by assigning a numerical score, this allows competing forecast distributions to be ranked. At every $t=n+h,\ldots,n+40$ these scores are based on the $h$-step-ahead posterior predictive distribution at time $t$, which is then averaged across the hold-out period. We chose two widely used scores to assess forecasting performance for the three variables of interest, individually: the continuous rank probability score (CRPS) and the logarithmic score. We additionally compare joint forecasts of the three variables of interest by computing the energy score (ES), which is a multivariate generalization of the CRPS. In all cases, the scores are negatively oriented so that small values indicate better forecasting performance. Further details on the calculation of the MSFE and the proper scoring rules, along with their adaptation for the maximum likelihood analysis, can be found in the Supplementary Materials. For each value of $m$ and each prior, the values of the one-step-ahead ($h=1$) CRPS, ES and posterior mean MSFE are shown in Figure~\ref{fig:scoring_rules}, along with approximations of the posterior probability that $\matr{\Phi} \in \mathcal{C}_{4,m}$. Analogous plots for the one-step-ahead logarithmic scores, which showed similar patterns to the CRPS, and for all metrics at the other horizons $h=2,4,8$ can be found in the Supplementary Materials.

For each value of $m$, and under all metrics, it is noticeable from Figure~\ref{fig:scoring_rules} that the exchangeable, stationary prior performs consistently well, especially when $m=20$, where it is ranked as best by all criteria. The Minnesota prior also forecasts well in this one-step-ahead setting. The frequentist forecasts appear successful in terms of the MSFE, but their performance is ranked worst by the proper scoring rules. \label{pg:rev1_5c}Since these forecasts ignore epistemic uncertainty in the parameter values, it is likely that this is because the forecast distributions are overly concentrated. The semi-conjugate prior and the vague, stationary prior perform poorly according to all metrics, which is likely because of the lack of structure imposed by modelling assumptions, in the former case, or encouraged by the prior, in the latter case, leading to very diffuse predictive distributions. Additional figures in the Supplementary Materials reveal how the performance of the various predictive distributions changes moving from one-quarter-ahead forecasts to up to two-year-ahead forecasts. Over longer time horizons $h$, the exchangeable, stationary prior is the only one whose performance is consistently strong according to all metrics, especially when $m=10$ and $m=20$ where differences to other forecasts are most apparent. The performance of the frequentist forecasts continue to appear successful in terms of the MSFE, but less so under the proper scoring rules. In contrast, the performance of the Minnesota prior deteriorates rapidly as $h$ increases, possibly due to its use of a fixed diagonal error variance matrix. As the forecast horizon increases, it is interesting to observe how the behaviour of the forecasts under the priors which do and do not enforce stationarity change with $m$. Under the stationarity-enforcing prior distributions, performance is generally similar, and sometimes better, going from $m=10$ to $m=20$. However, under the priors which do not enforce stationarity, there is a marked deterioration in forecasting performance. As $m$ increases from 10 to 20, the approximate posterior probability that $\matr{\Phi}$ lies inside the stationary region decreases from 0.8125 to 0.0000 under the Minnesota prior and from 0.8413 to 0.0000 under the semi-conjugate prior. At least in part, this is likely to be due to epistemic uncertainty spilling outside the stationary region, whose volume becomes vanishingly small when $p$ or, in this case, $m$ become large; see Section~\ref{sec:intro}. When the parameters of a \VAR{m}{p} model lie outside the stationary region, the forecast variance increases without bound into the future. Therefore one might reasonably conjecture that it is this greater concentration of posterior mass outside the stationary region that leads to the more noticeable deterioration in forecasting performance when inference is not constrained to the stationary region. This suggests that using a constrained prior distribution may be even more important in problems where $m$ or $p$ are large. Moreover, the better performance of the exchangeable, stationary prior over the vague, stationary prior illustrates the benefits of adopting a prior which encourages shrinkage towards sensible parametric structures.

\section{\label{sec:discussion}Discussion}

It is straightforward to extend the reparameterization and prior presented here to VARMA models. Consider the model of order $(p,q)$, or the \VARMA{m}{p}{q} model, $\matr{\theta}(B) \vec{\epsilon}_t = \matr{\phi}(B) \vec{y}_t$, where $\matr{\theta}(u) = \matr{I}_m + \matr{\theta}_1 u + \cdots + \matr{\theta}_q u^q$, $u \in \mathbb{C}$, is the characteristic moving average polynomial, in which $\matr{\theta}_i \in M_{m \times m}(\mathbb{R})$ ($i=1,\ldots,q$). As for \VAR{m}{p} models, the process is stationary if and only if all the roots of $\det \{ \matr{\phi}(u) \} = 0$ lie outside the unit circle. It is invertible if and only if all the roots of $\det \{ \matr{\theta}(u) \} = 0$ lie outside the unit circle, and hence the invertible region is $\mathcal{C}_{q,m}$. We can therefore constrain inference to the stationary and invertible regions simultaneously by reparameterizing the model in terms of two sets of matrices with singular values less than one. In other words, we can apply the recursions described in the Appendix twice; once as if we had a pure \VAR{m}{p} model with coefficients $\matr{\phi}_1, \ldots, \matr{\phi}_p$ and variance $\matr{\Sigma}$ to get $\matr{P}_1, \ldots, \matr{P}_p$, and again as if we had a pure \VAR{m}{q} model with coefficients $-\matr{\theta}_1, \ldots, -\matr{\theta}_q$ and variance $\matr{\Sigma}$ to get, say, $\matr{R}_1, \ldots, \matr{R}_q$. Unfortunately, interpretation of the new parameter sets is a little less clear; the $\matr{P}_s$ represent the partial autocorrelation matrices of the autoregressive part of the process, and the $\matr{R}_s$ represent a multivariate analogue of the inverse partial autocorrelation function \citep[][]{Bha83} for the moving average part of the process. In each case, the second transformation from Section~\ref{subsec:part2} can be used to map the parameters to unconstrained Euclidean space. Details on inference are given in the Supplementary Materials.

The ideas discussed in this paper can also be extended to VAR models of unknown order. This might be useful when there is little prior information to guide the choice of $p$. At least in theory, allowing for uncertainty in the model order is easy to handle in the Bayesian framework. Suppose we are prepared to consider models where the order does not exceed $p_{\text{max}}$. We can then make the model order, say $k$, unknown and assign it a prior over $\{ 0, \ldots, p_{\text{max}}\}$. As discussed in Section~\ref{subsec:part2}, the vector autoregression is of order $k < p_{\text{max}}$ if and only if $\matr{A}_k \ne \matrn{0}_m$ and $\matr{A}_{k+s} = \matrn{0}_m$ ($s=1,\ldots,p_{\text{max}}-k$) and so the models of different orders are nested. Various transdimensional MCMC samplers have been developed for problems like these, where the dimension of the parameter space is itself unknown. In the context of univariate autoregressions, this includes reversible jump MCMC \citep[][]{Gre95} and birth-death MCMC \citep[][]{Ste00}, with samplers that both do, and do not, enforce stationarity \citep[e.g.][]{VADG04,Phi06}. The advantages of tackling this problem with our reparameterization of the \VAR{m}{p} process is that it allows stationarity to be enforced whilst maintaining the nested structure of the models from the original parameterization. We defer further consideration of this interesting challenge to future work.

\section*{Acknowledgements}

This work was supported by the EPSRC grant EP/N510129/1 via the the Alan Turing Institute project ``Streaming data modelling for real-time monitoring and forecasting''. I would like to thank the referees and editors for their helpful comments and suggestions. I am also grateful to Gary Koop for providing access to the macroeconomic data and to Michael Betancourt, Malcolm Farrow, Tom Nye, Anindya Roy and Darren Wilkinson for conversations which have improved the manuscript.

\appendix

\section*{\label{app:mapping}Appendix}

\subsection*{Forward mapping: VAR parameters to partial autocorrelations}

This proceeds as described in \citet{AN79}. Using Cholesky factors in the matrix-square-roots below leads to the parameterization of \citet{AK86}, whilst using symmetric matrix-square-roots gives rise to the parameterization described in this paper.

The mapping from $( \matr{\Sigma}, \matr{\Phi} ) \in \mathcal{S}^+_m \times \mathcal{C}_{p,m}$ to $\{ \matr{\Sigma}, (\matr{P}_1, \ldots, \matr{P}_p) \} \in \mathcal{S}^+_m \times \mathcal{V}_{m}^p$, described in \citet{AN79}, proceeds in two main stages.

\begin{enumerate}[nosep,labelindent=0pt,leftmargin=*,label=\arabic*.,ref=\arabic*]
\item From $( \matr{\Sigma}, \matr{\Phi} )$, compute the autocovariances $\matr{\Gamma}_i = \Cov(\vec{y}_t, \vec{y}_{t+i})$ ($i=0, \ldots, p$). $\matr{\Gamma}_0, \ldots, \matr{\Gamma}_{p-1}$ can be found by representing the autoregression as a \VAR{pm}{1} process and computing its stationary variance. The resulting discrete Lyapunov equation can be solved using vectorization and Kronecker product operators. The remaining autocovariance $\matr{\Gamma}_p$ can then be calculated using the Yule-Walker equations for the order $p$ process. For further details, see, for example, Chapter 2 of \citet{Lut05}.
\item From $\matr{\Phi}$ and $(\matr{\Gamma}_0, \ldots, \matr{\Gamma}_p)$ compute $(\matr{P}_1, \ldots, \matr{P}_p)$:
\begin{enumerate}[nosep,labelindent=0pt,label=(\alph*),ref=\theenumi{}(\alph*)]
\item Initialize: construct $\matr{\Sigma}_0 = \matr{\Sigma}^*_0 = \matr{\Gamma}_0$ and then calculate the matrix-square-root factorizations, $\matr{\Sigma}_0 = \matr{\Sigma}^*_0 = \matr{S}_0 \matr{S}_0^\T = \matr{S}^*_0 \matr{S}^{* \, T}_0$.
\item Recursion: for $s=0,\ldots,p-1$:
\begin{enumerate}[nosep,labelindent=0pt,label=(\roman*),ref=\theenumii{}(\roman*)]
\item\label{eq:formap1and2} Compute $\matr{\phi}_{s+1,s+1}$ and $\matr{\phi}^*_{s+1,s+1}$ using
\begingroup
\allowdisplaybreaks
\begin{align*}
\matr{\phi}_{s+1,s+1} &= \left( \matr{\Gamma}_{s+1}^\T - \matr{\phi}_{s1} \matr{\Gamma}_s^\T - \cdots - \matr{\phi}_{ss} \matr{\Gamma}_1^\T \right) \matr{\Sigma}^{* \, -1}_s,\\
\matr{\phi}^*_{s+1,s+1} &= \left( \matr{\Gamma}_{s+1} - \matr{\phi}^*_{s1} \matr{\Gamma}_s - \cdots - \matr{\phi}^*_{ss} \matr{\Gamma}_1 \right) \matr{\Sigma}^{-1}_s,
\end{align*}
\endgroup
where it is understood that when $s=0$, these expressions simplify to $\matr{\phi}_{11} = \matr{\Gamma}_{1}^\T \matr{\Sigma}_0^{* \, -1}=\matr{\Gamma}_{1}^\T \matr{\Gamma}_0^{-1},\ \matr{\phi}_{11}^* = \matr{\Gamma}_{1} \matr{\Gamma}_0^{-1}$.
\item\label{eq:formap3and4} If $s>0$, for $i=1,\ldots,s$, compute $\matr{\phi}_{s+1,i}$ and $\matr{\phi}^*_{s+1,i}$ using
\begin{equation*}
\matr{\phi}_{s+1,i} = \matr{\phi}_{si} - \matr{\phi}_{s+1,s+1} \matr{\phi}^*_{s,s-i+1},\quad
\matr{\phi}^*_{s+1,i} = \matr{\phi}^*_{si} - \matr{\phi}^*_{s+1,s+1} \matr{\phi}_{s,s-i+1}.
\end{equation*}
\item\label{eq:formap5and6} Compute $\matr{P}_{s+1}$ using one of
\begin{equation*}
\matr{P}_{s+1} = \matr{S}_s^{-1} \matr{\phi}_{s+1,s+1} \matr{S}^*_s, \quad \matr{P}_{s+1} = \left(\matr{S}^{* \, -1}_s \matr{\phi}^*_{s+1,s+1} \matr{S}_s\right)^\T.
\end{equation*}
\item\label{eq:formap7and8} If $s < p-1$, compute $\matr{\Sigma}_{s+1}$ and $\matr{\Sigma}^*_{s+1}$ using 
\begin{align*}
\begin{split}
\matr{\Sigma}_{s+1} &= \matr{\Gamma}_0 - \matr{\phi}_{s+1,1} \matr{\Gamma}_1 - \cdots - \matr{\phi}_{s+1,s+1} \matr{\Gamma}_{s+1},\\
\matr{\Sigma}^*_{s+1} &= \matr{\Gamma}_0 - \matr{\phi}^*_{s+1,1} \matr{\Gamma}_1^\T - \cdots - \matr{\phi}^*_{s+1,s+1} \matr{\Gamma}_{s+1}^\T,
\end{split}
\end{align*}
then calculate the matrix-square-roots, $\matr{\Sigma}_{s+1} = \matr{S}_{s+1} \matr{S}_{s+1}^\T,\ \matr{\Sigma}^*_{s+1} = \matr{S}^*_{s+1} \matr{S}^{* \, T}_{s+1}$.
\end{enumerate}
\end{enumerate}
\end{enumerate}

\subsection*{Reverse mapping: partial autocorrelations to VAR parameters} 

The inverse mapping from $\{ \matr{\Sigma}, (\matr{P}_1, \ldots, \matr{P}_p) \} \in \mathcal{S}^+_m \times \mathcal{V}_{m}^p$ to $( \matr{\Sigma}, \matr{\Phi} ) \in \mathcal{S}^+_m \times \mathcal{C}_{p,m}$, comprises two recursions; the first is new and the second is based on Lemma 2.1 of \citet{AK86}.

\begin{enumerate}[nosep,labelindent=0pt,leftmargin=*,label=\arabic*.,ref=\arabic*]
\item From $\{ \matr{\Sigma}, (\matr{P}_1, \ldots, \matr{P}_p) \}$ compute the stationary variance matrix $\matr{\Gamma}_0$:
\begin{enumerate}[nosep,labelindent=0pt,label=(\alph*),ref=\theenumi{}(\alph*)]
\item Initialize: let $\matr{\Sigma}_p = \matr{\Sigma}$ with corresponding matrix-square-root factorization, $\matr{\Sigma}_p = \matr{S}_p \matr{S}_p^\T$.
\item\label{eq:revmap1} Recursion: for $s=p-1,\ldots,0$ construct the symmetric (or lower triangular) matrix $\matr{S}_s$ such that
\begin{equation*}
\matr{\Sigma}_{s+1} = \matr{S}_s (\matr{I}_m - \matr{P}_{s+1} \matr{P}_{s+1}^\T) \matr{S}_s^\T,
\end{equation*}
then compute $\matr{\Sigma}_s = \matr{S}_s \matr{S}_s^\T$.
\item Output: take $\matr{\Gamma}_0 = \matr{\Sigma}_{0}$.
\end{enumerate}
\item From $(\matr{P}_1, \ldots, \matr{P}_p)$ and $\matr{\Gamma}_0$ compute the matrices in $\matr{\Phi}$:
\begin{enumerate}[nosep,labelindent=0pt,label=(\alph*),ref=\theenumi{}(\alph*)]
\item Initialize: let $\matr{\Sigma}_0 = \matr{\Sigma}^*_0 = \matr{\Gamma}_0$ with corresponding matrix-square-root factorization, $\matr{\Sigma}_0 = \matr{\Sigma}^*_0 = \matr{S}_0 \matr{S}_0^\T = \matr{S}^*_0 \matr{S}^{* \, T}_0$.
\item Recursion: for $s=0,\ldots,p-1$:
\begin{enumerate}[nosep,labelindent=0pt,label=(\roman*),ref=\theenumii{}(\roman*)]
\item\label{eq:revmap2and3} Compute $\matr{\phi}_{s+1,s+1}$ and $\matr{\phi}^*_{s+1,s+1}$ using
\begin{equation*}
\matr{\phi}_{s+1,s+1} = \matr{S}_s \matr{P}_{s+1} \matr{S}^{* \, -1}_s, \quad \matr{\phi}^*_{s+1,s+1} = \matr{S}^*_s \matr{P}_{s+1}^\T \matr{S}^{-1}_s.
\end{equation*}
\item\label{eq:revmap4and5} If $s>0$, for $i=1,\ldots,s$, compute $\matr{\phi}_{s+1,i}$ and $\matr{\phi}^*_{s+1,i}$ using
\begin{equation*}
\matr{\phi}_{s+1,i} = \matr{\phi}_{si} - \matr{\phi}_{s+1,s+1} \matr{\phi}^*_{s,s-i+1}, \quad \matr{\phi}^*_{s+1,i} = \matr{\phi}^*_{si} - \matr{\phi}^*_{s+1,s+1} \matr{\phi}_{s,s-i+1}.
\end{equation*}
\item\label{eq:revmap6and7} Compute $\matr{\Sigma}_{s+1}$ and $\matr{\Sigma}^*_{s+1}$ using
\begin{equation*}
\matr{\Sigma}_{s+1} = \matr{\Sigma}_s - \matr{\phi}_{s+1,s+1} \matr{\Sigma}^*_s \matr{\phi}^\T_{s+1,s+1}, \quad \matr{\Sigma}^*_{s+1} = \matr{\Sigma}^*_s - \matr{\phi}^*_{s+1,s+1} \matr{\Sigma}_s \matr{\phi}^{* \, T}_{s+1,s+1},
\end{equation*}
then calculate the matrix-square-roots, $\matr{\Sigma}_{s+1} = \matr{S}_{s+1} \matr{S}_{s+1}^\T$ and $\matr{\Sigma}^*_{s+1} = \matr{S}^*_{s+1} \matr{S}^{* \, T}_{s+1}$.
\item\label{eq:revmap8} Compute $\matr{\Gamma}_{s+1}$ using
\begin{equation*}
\matr{\Gamma}_{s+1}^\T = \matr{\phi}_{s+1,s+1} \matr{\Sigma}_s^* + \matr{\phi}_{s1} \matr{\Gamma}_s^\T + \cdots + \matr{\phi}_{ss} \matr{\Gamma}_1^\T.
\end{equation*}
\end{enumerate}
\item Output: take $\matr{\phi}_i = \matr{\phi}_{pi}$ ($i=1,\ldots,p$). By construction, $\matr{\Sigma} = \matr{\Sigma}_p$.
\end{enumerate}
\end{enumerate}

\bibliography{journal_names_abbr,refs}

\end{document}